\documentclass{PoS}

\usepackage{amsmath}
\usepackage{amsthm}
\usepackage{amssymb}
\usepackage{subcaption}

\renewcommand{\>}{\rangle}
\newcommand{\tr}{\text{tr}}
\renewcommand{\d}{\text{d}}
\newcommand{\e}{\text{e}}
\newcommand{\ddx}{\frac{\d}{\d x}}

\newcommand{\sch}[1]{}

\title{Quantum Critical Phenomena in an $O(4)$ Fermion Chain}

\ShortTitle{Quantum Critical Phenomena in an $O(4)$ Fermion Chain}

\author{\speaker{Hanqing Liu}\thanks{Work done in collaboration with Shailesh Chandrasekharan and Ribhu Kaul}\\
Department of Physics, Box 90305, Duke University, Durham, NC 27708, USA\\
E-mail: \email{hanqing.liu@duke.edu}}

\abstract{We construct a fermionic lattice model containing interacting spin-$\frac{1}{2}$ fermions with an $O(4)$ symmetry. In addition the model contains a $\mathbb{Z}_2$ chiral symmetry which prevents a fermion mass term. Our model is motivated by the ability to study its physics using the meron-cluster algorithm. By adding a strong repulsive Hubbard interaction $U$, we can transform it into the regular Heisenberg anti-ferromagnet. While we can study our model in any dimension, as a first project we study it in one spatial dimension. We discover that our model at $U=0$ can be described as a lattice-regularized 2-flavor Gross-Neveu model, where fermions become massive since the $\mathbb{Z}_2$ chiral symmetry of the model is spontaneously broken. We show numerically that the theory remains massive when $U$ is small. At large values of $U$ the model is equivalent to the isotropic spin-half anti-ferromagnetic chain, which is massless for topological reasons. This implies that our model has a quantum phase transition from a $\mathbb{Z}_2$ broken massive phase to a topologically massless phase as we increase $U$. We present results obtained from our quantum Monte Carlo method near this phase transition.}

\FullConference{37th International Symposium on Lattice Field Theory - Lattice2019\\
		16-22 June 2019\\
		Wuhan, China}

\begin{document}

\section{Introduction}
It has long been known that many nonlinear sigma models on homogeneous target spaces with topological terms can be studied through various quantum spin systems at their critical points both in $1+1$d \cite{Haldane:1982rj, Affleck:1987ch, Affleck:1988wz, Beard:2004jr, Affleck:1984ar} and $2+1$d \cite{Senthil:2004aza, Senthil:2005jk}. In this contribution, we will study the possibility of inducing a topological term in a lattice-regularized Gross-Neveu model \cite{Gross:1974jv} by adding a Hubbard interaction.
We first introduce a lattice model containing interacting spin-$\frac{1}{2}$ fermions with an $O(4)$ symmetry and a $\mathbb{Z}_2$ chiral symmetry. Since in the continuum close to the Gaussian fixed point there is a unique marginal coupling respecting these symmetries, we believe our model can be viewed as a lattice regularization of that continuum model, which is the usual Gross-Neveu model. Perturbative continuum analysis suggests that the marginal coupling can be relevant or irrelevant depending on the sign of the coupling. When the coupling is relevant one gets the usual massive phase of the Gross-Neveu model. Here we provide evidence that our lattice model is in this phase. Adding a strong Hubbard interaction to our model, we can argue that it becomes the quantum spin-half Heisenberg chain, whose low energy physics can be described by the $SU(2)_1$ Wess-Zumino-Witten (WZW) model \cite{Witten:1983ar}, with an additional marginally irrelevant operator. Thus, by tuning the Hubbard interaction $U$ slowly away from zero, we can study the phase transition between the massive phase of the Gross-Neveu model and the massless phase of the spin-half Heisenberg chain. Using the meron-cluster algorithm we discover that this transition is second order and at the quantum critical point the marginally irrelevant operator present in the spin-half Heisenberg chain seems to disappear. We also provide evidence for the enlarged symmetry corresponding to the WZW model \cite{tsvelik_2003}. 

This contribution is organized in the following way. In Section \ref{sec:Lattice}, we will introduce our lattice Hamiltonian and show the relevant symmetries. In Section \ref{sec:continuum}, we connect our model to the continuum Gross-Neveu model based on symmetry arguments, and discuss the quantum phase transition from the point of view of RG flows. Finally in Section \ref{sec:numerical}, we show some numerical results to support our theoretical analysis.

\section{The Lattice Model and its Symmetry}\label{sec:Lattice}
We work in the imaginary time Hamiltonian formalism. The theory is defined by the partition function $Z = \tr~ \e^{-\beta H}$, where the Hamiltonian is given by
\begin{align}
  H = H_J + H_U = -J  \sum_{\langle i,j\rangle}H_{\langle i,j\rangle\uparrow}H_{\langle i,j\rangle\downarrow} + U\sum _{i}\left({n}_{i\uparrow }-\frac{1}{2}\right)\left({n}_{i\downarrow }-\frac{1}{2}\right).
\label{eq:ourmodel}
\end{align}
The first term $H_J$ is a nearest neighbor term motivated by the ability to solve the physics using the meron-cluster algorithm, where
\begin{align}
 H_{\langle i,j\rangle\alpha} = -({c}_{i\alpha }^{\dagger }{c}_{j\alpha }+{c}_{j\alpha }^{\dagger }{c}_{i\alpha }) + 2\left({n}_{i\alpha }-\frac{1}{2}\right)\left({n}_{j\alpha}-\frac{1}{2}\right)-\frac{1}{2}.
\end{align}
The symbol $\langle i,j\rangle$ refers to that $i$ and $j$ are assumed to be nearest neighbors. The second term $H_U$ is the usual on-site Hubbard interaction between spin-up and spin-down fermions.

The symmetries of both terms are manifest in terms of Majorana operators, where we define two Majorana operators $\gamma_j^1$ and $\gamma_j^2$ for spin-up fermions on each lattice site $j$ through the relations
\begin{align}
c_{j\uparrow}&=\frac{1}{2}(\gamma_j^{1}-i\gamma_j^{2}), \quad c^\dagger_{j\uparrow}=\frac{1}{2}(\gamma_j^{1}+i\gamma_j^{2}), \quad n_{j\uparrow}=\frac{1}{2} (-i\gamma_j^{1}\gamma_j^{2}+1), \text{ for $j$ even,} \nonumber \\
c_{j\uparrow}&=\frac{1}{2}(\gamma_j^2+i\gamma_j^1), \quad c^\dagger_{j\uparrow}=\frac{1}{2}(\gamma_j^2-i\gamma_j^1), \quad n_{j\uparrow}=\frac{1}{2} (-i\gamma_j^1\gamma_j^2+1), \text{ for $j$ odd.}
\end{align}
Similarly we can define two more Majorana operators $\gamma_j^3$ and $\gamma_j^4$ using the spin-down fermions. In terms of the Majorana operators the Hamiltonian takes the form
\begin{align}
  H = H_J + H_U = -\frac{J}{4}\sum_{\langle i,j\rangle}\prod_{\mu=1}^4(1+i\gamma_{i}^\mu\gamma_{j}^\mu) -\frac{U}{96}\sum_i \varepsilon_{\mu\nu\rho\sigma}\gamma_i^\mu \gamma_i^\nu \gamma_i^\rho \gamma_i^\sigma. \nonumber
\end{align}
Note that $[H, \Gamma^{\mu\nu}] = 0$, where $\Gamma^{\mu\nu} = i\sum_{j}\gamma_j^\mu\gamma_j^\nu$ satisfy the $\mathfrak{so}(4)$ algebra.
The algebra $\mathfrak{so}(4)$ is isomorphic to $\mathfrak{su}(2)_s \times \mathfrak{su}(2)_c$, which corresponds to the spin and charge symmetries respectively as explained in the next section. There is an additional $\mathbb{Z}_2$ transformation $P = i\sum_{j}\gamma_j^{1}\gamma_j^{3}\gamma_j^{4}$ which corresponds to a spin-charge flip, also as explained in the next section.
Since $PH_JP = H_J$ and $PH_UP = -H_U$, $H_J$ has $O(4)$ symmetry, while $H_U$ only has $SO(4)$ symmetry. In addition, the lattice model is invariant under $T_a$, i.e. translation by one lattice site, which is a discrete subgroup of the full continuum chiral symmetry as discussed below.

\section{Connection to the Continuum Gross-Neveu Model}
\label{sec:continuum}
In this section we argue that our lattice model $H_J$ in Eq.(\ref{eq:ourmodel}) is related to the continuum Gross-Neveu model. If we ignore all the interactions and focus on the hopping part of the Hamiltonian $H$, at low energies we obtain the free continuum Dirac fermion theory described by,
\begin{align}
    H_0^\text{cont} = \sum_{\alpha=1,2}\int \d x\left(-\psi^\dagger_{\alpha,L}(x)i\ddx\psi_{\alpha,L}(x)+\psi^\dagger_{\alpha,R}(x)i\ddx\psi_{\alpha,R}(x)\right).
\end{align}
This Hamiltonian is invariant under $SU(2)_{s,L}\times SU(2)_{c,L} \times SU(2)_{s,R}\times SU(2)_{c,R}$, which we refer to as the full chiral symmetry. The $SU(2)_s\times SU(2)_c$ symmetry in each chiral sector can be made manifest by arranging the fermion fields into two $2\times 2$ matrix-valued fields,
\begin{align}
  \Psi_{L, R}(x) = \begin{pmatrix}
    \psi_{L, R}^1(x) & \psi_{L, R}^{2\dagger}(x) \\
    \psi_{L, R}^2(x) & -\psi_{L, R}^{1\dagger}(x)
  \end{pmatrix},
\end{align}
where each column (row) transforms as a spinor under $SU(2)_{s(c)}$ in that sector. We can then rewrite
\begin{align}
    H_0^\text{cont} \ =\ \int \d x \ \frac{1}{2}\tr\left(-\Psi_L^\dag(x) i\ddx \Psi_L(x) \ +\ \Psi_R^\dag(x) i\ddx \Psi_R(x)\right),
\end{align}
whose chiral $SU(2)_s\times SU(2)_c$ symmetry can be read out immediately: $SU(2)_{s(c)}$ acts as left (right) multiplication on $\Psi_L(x)$ or $\Psi_R(x)$ independently. Note that $H_0^\text{cont}$ is also invariant under the spin-charge flip symmetries $P_{L,R}: \Psi_{L,R}(x) \rightarrow \Psi_{L,R}^\dagger(x)$ in each chiral sector.

Let us now consider the possible interactions which respect the symmetries on the lattice. This requires us to understand how the symmetries on the lattice are translated into the continuum. The $SO(4)$ symmetry on the lattice is mapped to the diagnonal part of the full chiral symmetry. The spin-charge flip is $P = P_L + P_R$, while the translation symmetry $T_a$ is the $\mathbb{Z}_2$ chiral transformation $\psi_{L}^{\alpha}(x) \rightarrow i\psi_{L}^{\alpha}(x)$ and $\psi_{R}^{\alpha}(x) \rightarrow -i\psi_{R}^{\alpha}(x)$
which belongs to $SU(2)_{cL} \times SU(2)_{cR}$.

Near the free fermion (Gaussian) fixed point, quadratic terms are relevant, quartic terms are marginal and higher terms are irrelevant. We will focus on the first two cases. There are no quadratic terms respecting these lattice symmetries. As for the quartic terms, it was argued by Affleck \cite{Affleck:1988zj} that terms preserving the full chiral symmetry do not change the physics. Thus, perturbatively the continuum theory can only contain qaurtic terms that preserve the diagonal $SO(4)$ as well as $T_a$. The only two possible terms are $S_L^iS_R^i$ and $Q_L^iQ_R^i$, where 
\begin{align}
  S^i_{L, R}(x) =& \frac{1}{2}\tr ~\Psi_{L, R}^\dag \sigma^i \Psi_{L, R} = \psi_{L, R}^{\alpha\dagger} \sigma^i_{\alpha \beta}\psi_{L, R}^{\beta}, \\
  Q^{i}_{L, R}(x) =& \frac{1}{2}\tr ~\Psi_{L, R} \sigma^i \Psi_{L, R}^\dag =
                   \begin{pmatrix}
                     \psi_{L, R}^1 & \psi_{L, R}^{2\dagger}
                   \end{pmatrix}
                                 \sigma^i
                                 \begin{pmatrix}
                                   \psi_{L, R}^{1\dagger} \\
                                   \psi_{L, R}^2
                                 \end{pmatrix}
\end{align}
are spin and charge densities. They rotate as vectors under the left or right $SU(2)_s$ and $SU(2)_c$ respectively. Hence our lattice model must be related to the following continuum Hamiltonian
\begin{align}
  H^\text{cont}(\lambda_s,\lambda_c) &= \int \d x \sum_{\alpha} \left(-\psi^\dagger_{\alpha,L}(x)i\ddx\psi_{\alpha,L}(x)+\psi^\dagger_{\alpha,R}(x)i\ddx\psi_{\alpha,R}(x)\right) + \lambda_sS_L^iS_R^i + \lambda_cQ_L^iQ_R^i.
\end{align}
Since $H_J$ is also invariant under the $P$ (spin-charge flip) symmetry, we must have $\lambda_s = \lambda_c =:\lambda$ for $H_J$. We can then rewrite the interaction as
\begin{align}
  \lambda \left(S_L^iS_R^i + Q_L^iQ_R^i\right) = \lambda \left(M^2 + 1\right),
\end{align}
where $M = i~\tr\Psi_{L}^\dag \Psi_{R} = i\left(\psi_{L\alpha}^\dagger \psi_R^\alpha - \psi_{R\alpha}^\dagger \psi_L^\alpha\right)$ preserves the diagonal $O(4)$ symmetry manifestly.
Note that under a chiral field redefinition $\psi_{L\alpha} \rightarrow e^{i\frac{\pi}{4}}\psi_{L\alpha}$ and $\psi_{R\alpha} \rightarrow e^{-i\frac{\pi}{4}}\psi_{R\alpha}$, we have $M = \psi_{L\alpha}^\dagger \psi_R^\alpha + \psi_{R\alpha}^\dagger \psi_L^\alpha$, which is the usual mass term. Therefore we recognize $H^\text{cont}(\lambda,\lambda)$ as the two flavor Gross-Neveu model, which has the following $\beta$ function \cite{Gross:1974jv},
\begin{align}
  \frac{\d\lambda}{\d\log\mu} = -\frac{\lambda^2}{2\pi}.
\end{align}
When $\lambda > 0$, the model is asymptotically free and a mass scale is generated dynamically. As a result, the $\mathbb{Z}_2$ chiral symmetry $T_a$, which flips the sign of the mass term, is spontaneously broken and fermions become massive. All this suggests that the model $H_J$ is a lattice regularization of the Gross-Neveu model and numerical evidence shows we obtain a massive phase.

\begin{figure}[t]
     \centering \includegraphics[width=0.4\textwidth]{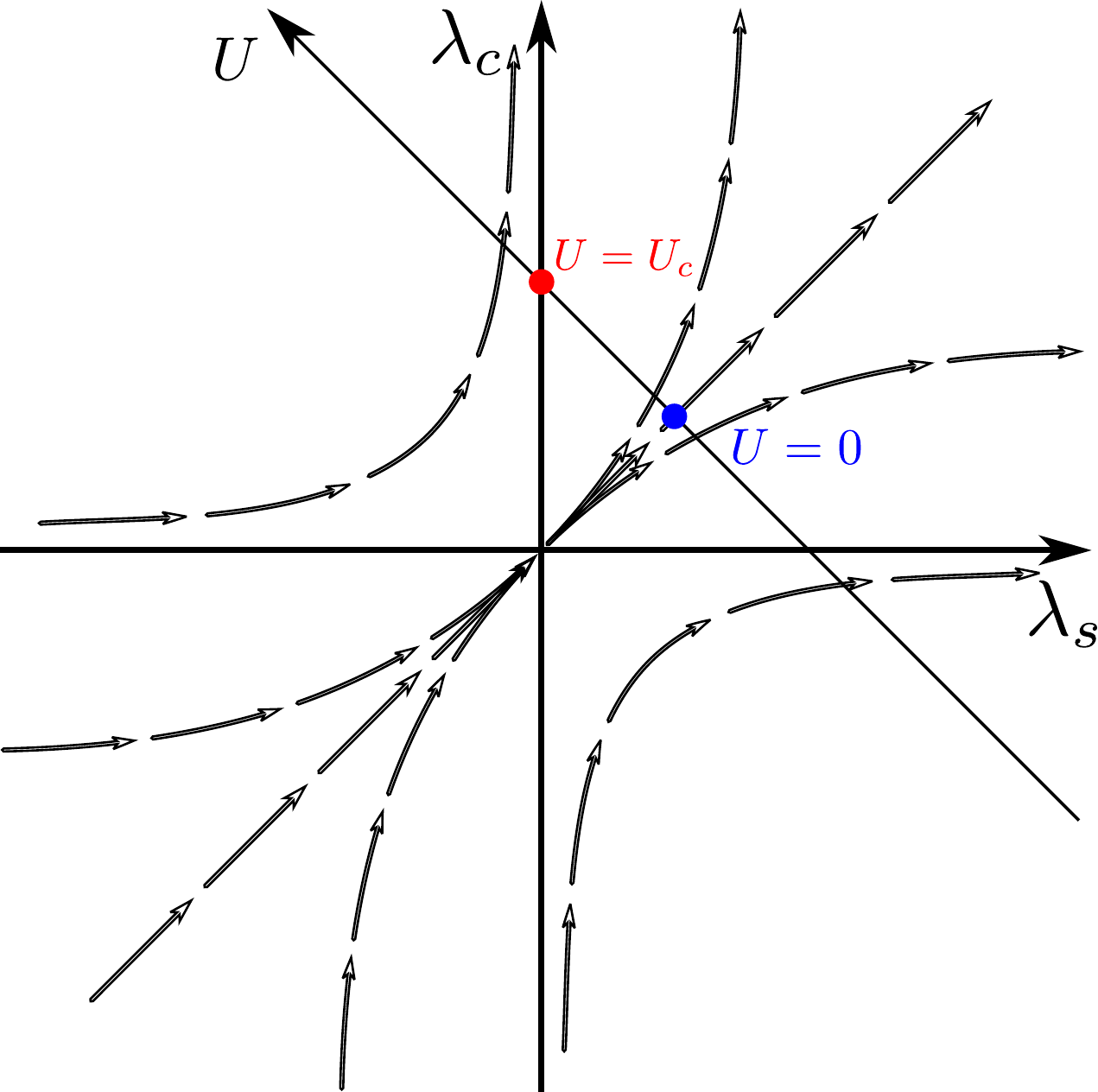}
  \caption{Flow diagram in the $\lambda_s - \lambda_c$ plane. \sch{Can you make the text not be covered by lines?}}
  \label{fig:flow-diagram}
\end{figure}

Since we know that the lattice Hamiltonian $H_U$ is odd under the spin-charge flip $P$, we conclude that near the Gaussian fixed point the corresponding continuum Hamiltonian take the form
\begin{align}
  H_U^\text{cont} &= U'\int \d x \left(-S_L^iS_R^i + Q_L^iQ_R^i\right).
\end{align}
where $U'$ is related to $U$ in some way. 
Thus by tuning $U$ we are able to explore the physics of $H^\text{cont}(\lambda_s,\lambda_c)$ away from the symmetric point $\lambda_s = \lambda_c$. In perturbation theory the couplings $\lambda_{s,c}$ satisfy independent $\beta$ functions \cite{Affleck:1988zj},
\begin{align}
  \frac{\d\lambda_s}{\d\log\mu} = -\frac{\lambda_s^2}{2\pi}, \quad \frac{\d\lambda_c}{\d\log\mu} = -\frac{\lambda_c^2}{2\pi},
\end{align}
and the flow diagram is shown in Fig.\ref{fig:flow-diagram}. We also schematically show the line of coupling $U$ in this space.

\section{Quantum Monte Carlo Results}
\label{sec:numerical}
\begin{figure}[h]
  \begin{subfigure}{0.5\textwidth}
    \includegraphics[width=1\textwidth]{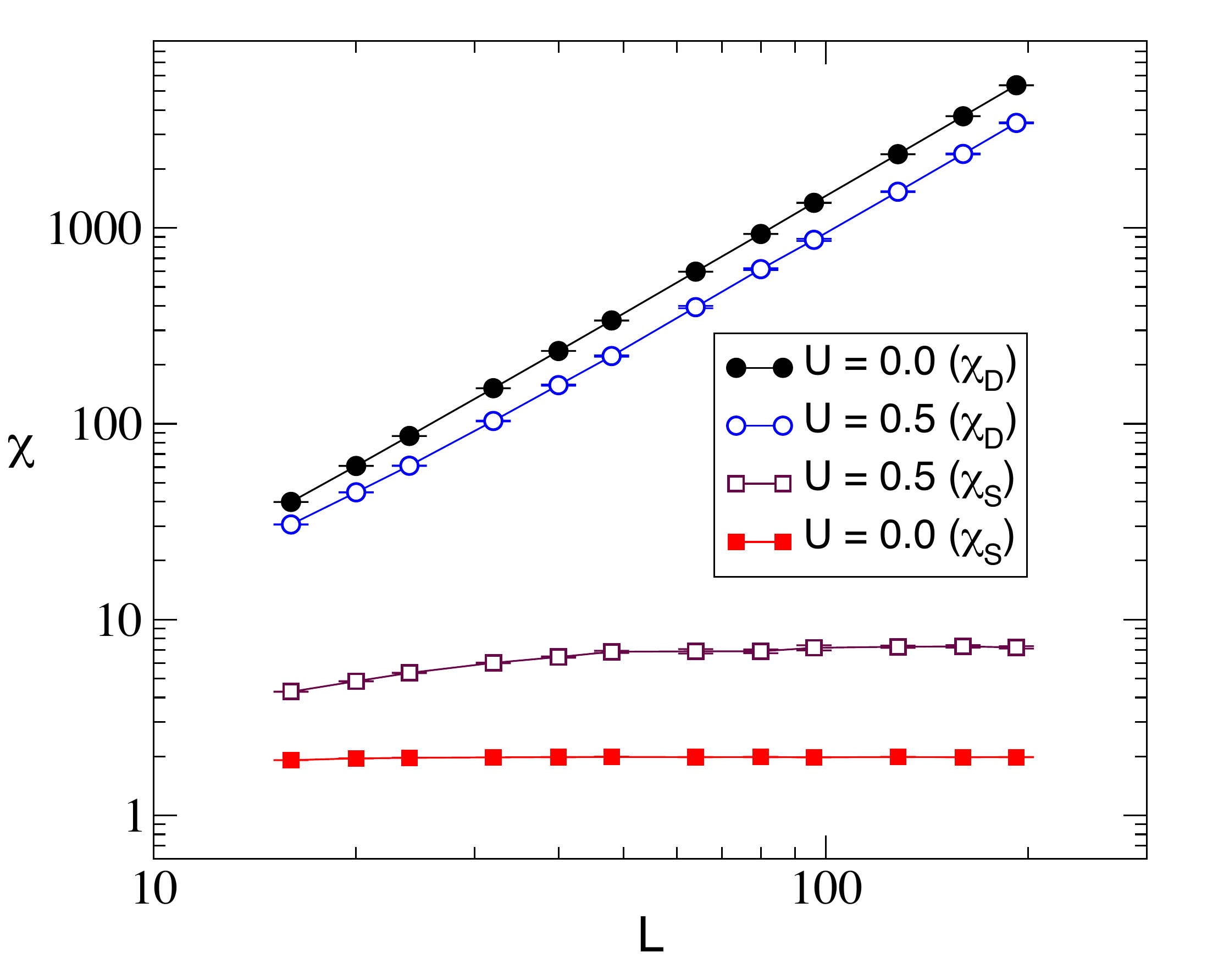}
    \caption{$\chi_S$ and $\chi_D$ as a function of $L$ at $U < U_c$.}
    \label{fig:S-L}
  \end{subfigure}
  \begin{subfigure}{0.49\textwidth}
    \includegraphics[width=1\textwidth]{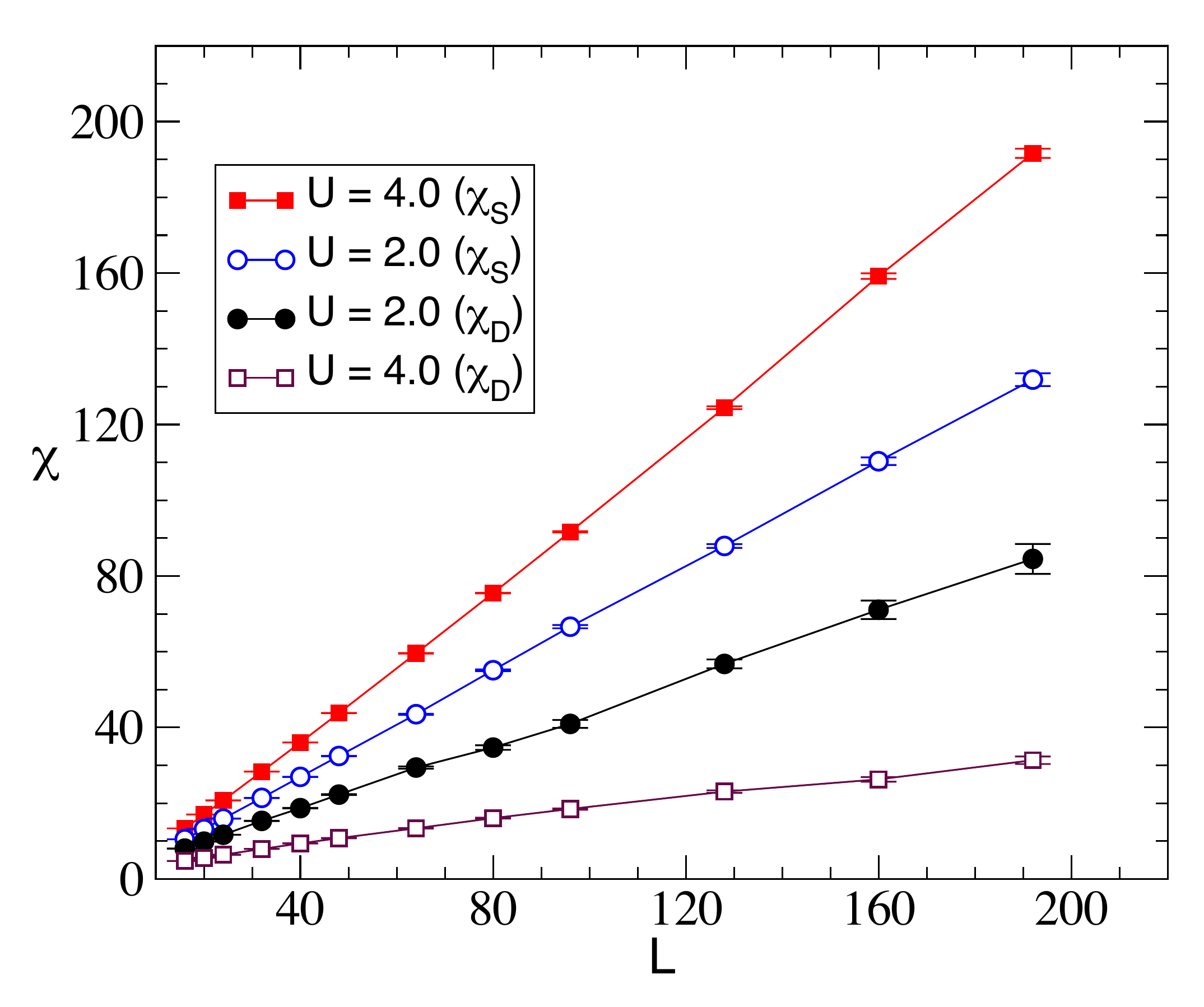}
    \caption{$\chi_S$ and $\chi_D$ as a function of $L$ at $U > U_c$.}
    \label{fig:D-L}
  \end{subfigure}
  \caption{Below $U_c$, $\chi_S$ saturates while $\chi_D$ diverges as $L^2$; above $U_c$ both susceptibilities grow linearly with log corrections.}
  \label{fig:L}
\end{figure}

We have studied our model in continuous time formalism by extending the meron-cluster algorithm developed earlier in \cite{Chandrasekharan:1999cm, Chandrasekharan:2002vk}. Details of the algorithm will be presented in a forthcoming publication \cite{Liu:2019}. The configurations we sample contain information about fermion occupation numbers $n_{j\alpha}(t)$ in space-time. In addition every configuration is described by loop clusters which provide information about correlations among the fermions. Using this information we study correlation functions of spins $S_j^z(t) = \frac{1}{2}(n_{j\uparrow}-n_{j\downarrow})(t)$ at lattice site $j$ and time $t$. We also study correlation functions of dimers $D_j(t) = \frac{1}{2}(S^z_jS^z_{j+1}-S^z_{j-1}S^z_j)(t)$.
In particular we have measured the spin and dimer susceptibilities $\chi_S$ and $\chi_D$ defined as
\begin{align}
  \chi_S \ &=\ \int dt\ \sum_{j}\ (-1)^j\<S^z_0(0) S^z_j(t)\>,\\
  \chi_D \ &=\ \int dt\ \sum_{j}\ (-1)^j\<D_0(0)D_j(t)\>.
\end{align}
In Fig.\ref{fig:L} we plot the lattice size dependence of these susceptibilities at $U=0,0.5,2,4$. We find that when $U<U_c\approx 1.7$ (Fig.\ref{fig:S-L}), $\chi_S$ saturates while $\chi_D$ grows as $L^2$ suggesting that we are in a phase where fluctuations of the spins are massive but there is long range order in the dimer order parameter $D_j(t)$. This is consistent with the broken $\mathbb{Z}_2$ chiral symmetry $T_a$. On the other hand, when $U>U_c$ (Fig.\ref{fig:D-L}) we find that both susceptibilities increase linearly (plus logarithmic corrections) with $L$. This is consistent with the model being in the WZW conformal phase similar to the spin-half Heisenberg chain, which is recovered in the $U=\infty$ limit.

\begin{figure}[h]
  \begin{subfigure}{0.5\textwidth}
    \includegraphics[width=1\textwidth]{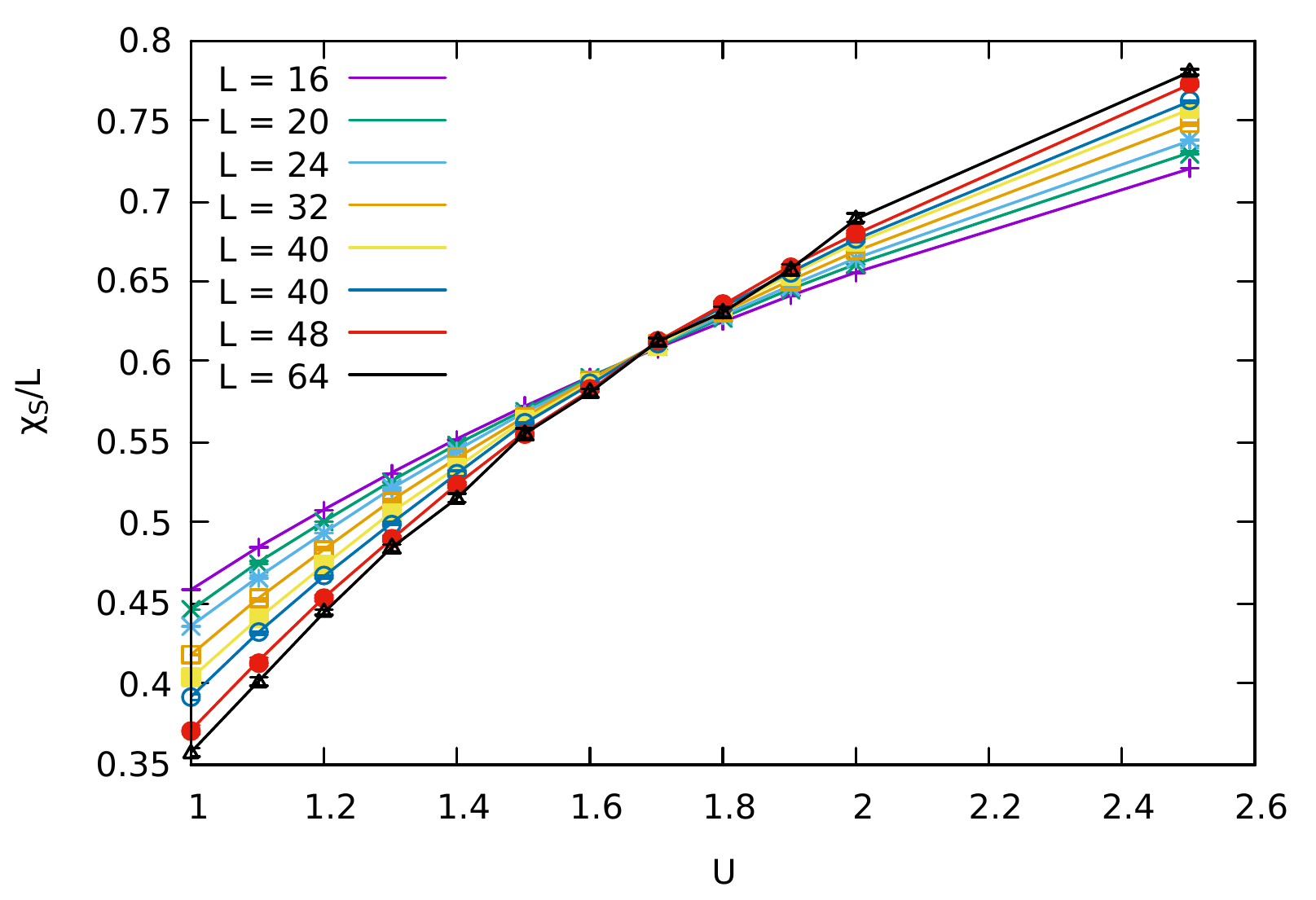}
    \caption{$\chi_S/L$ as a function of $U$}
    \label{fig:S-U}
  \end{subfigure}
  \begin{subfigure}{0.5\textwidth}
    \includegraphics[width=1\textwidth]{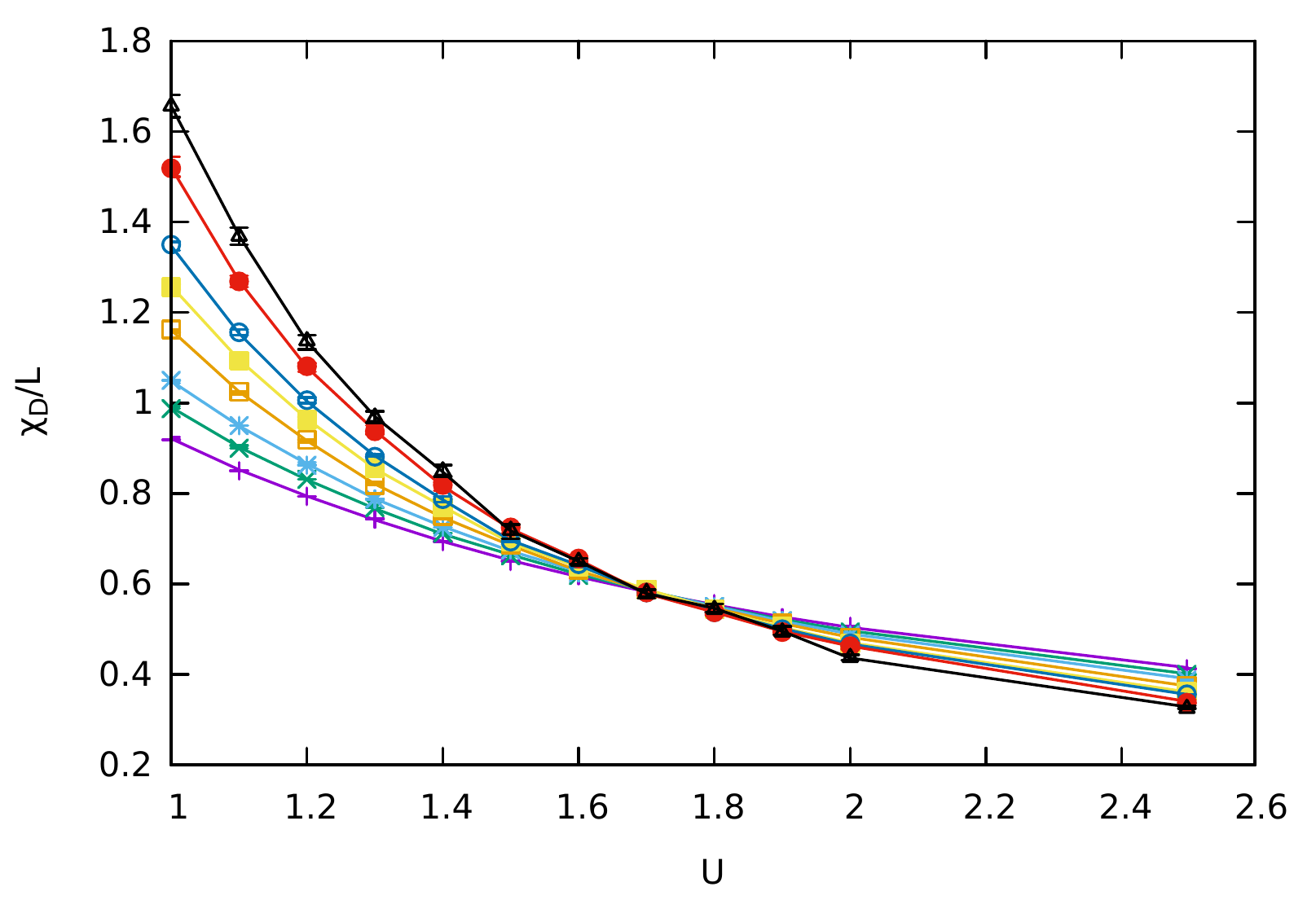}
    \caption{$\chi_D/L$ as a function of $U$}
    \label{fig:D-U}
  \end{subfigure}
  \caption{$\chi_S/L$ and $\chi_D/L$ as a function of $U$ at various $L$ cross at $U_c\approx 1.7$.}
  \label{fig:U}
\end{figure}

Based on Fig.\ref{fig:flow-diagram} we see that at the critical point $U = U_c$, we expect $\lambda_s=0$ which means the marginal operator $S_L^iS_R^i$ vanishes and the low energy theory is described exactly by the $SU(2)_1$ WZW model. The spin operators $\vec{S}(x,t) \approx \vec{S}_i(t)$ and the dimer operator $D(x,t) \approx D_i(t)$ combine and form a larger $SO(4)$ vector describing the $SU(2)$ WZW field, $g(x,t) = D(x,t) + i \vec{S}(x,t)\cdot \vec{\sigma}$ \cite{tsvelik_2003}. Thus we expect that at $U=U_c$ spin and dimer fields have the same scaling dimension, which is known to be $1/2$ from this bosonic field theory. This implies $\chi_S$ and $\chi_D$ both scale linearly in $L$. In Fig.\ref{fig:U} we plot $\chi_S/L$ and $\chi_D/L$ as a function of $U$ at various $L$ and they cross at $U = U_c$. We find that at $U_c\approx 1.7$, $\chi_S \sim L^{0.9994(18)}$ and $\chi_D \sim L^{1.0028(72)}$. 

\section{Acknowldegments}

The author would like to thank Shailesh Chandrasekharan for carefully reviewing and revising this contribution. The material presented here is supported by the U.S. Department of Energy, Office of Science, Nuclear Physics program under Award Numbers DE-FG02-05ER41368.

\bibliographystyle{JHEP}
\bibliography{Refs}

\providecommand{\href}[2]{#2}\begingroup\raggedright\begin{thebibliography}{10}

\bibitem{Haldane:1982rj}
F.~D.~M. Haldane, \emph{{Continuum dynamics of the 1-D Heisenberg
  antiferromagnetic identification with the O(3) nonlinear sigma model}},
  \href{http://dx.doi.org/10.1016/0375-9601(83)90631-X}{\emph{Phys. Lett.} {\bf
  A93} (1983) 464--468}.

\bibitem{Affleck:1987ch}
I.~Affleck and F.~D.~M. Haldane, \emph{{Critical Theory of Quantum Spin
  Chains}}, \href{http://dx.doi.org/10.1103/PhysRevB.36.5291}{\emph{Phys. Rev.}
  {\bf B36} (1987) 5291--5300}.

\bibitem{Affleck:1988wz}
I.~Affleck, \emph{{Critical Behavior of SU($n$) Quantum Chains and Topological
  Nonlinear $\sigma$ Models}},
  \href{http://dx.doi.org/10.1016/0550-3213(88)90117-4}{\emph{Nucl. Phys.} {\bf
  B305} (1988) 582--596}.

\bibitem{Beard:2004jr}
B.~B. Beard, M.~Pepe, S.~Riederer and U.~J. Wiese, \emph{{Study of CP(N-1)
  theta-vacua by cluster-simulation of SU(N) quantum spin ladders}},
  \href{http://dx.doi.org/10.1103/PhysRevLett.94.010603}{\emph{Phys. Rev.
  Lett.} {\bf 94} (2005) 010603},
  [\href{http://arxiv.org/abs/hep-lat/0406040}{{\tt hep-lat/0406040}}].

\bibitem{Affleck:1984ar}
I.~Affleck, \emph{{The Quantum Hall Effect, $\sigma$ Models at $\theta = \pi$
  and Quantum Spin Chains}},
  \href{http://dx.doi.org/10.1016/0550-3213(85)90353-0}{\emph{Nucl. Phys.} {\bf
  B257} (1985) 397--406}.

\bibitem{Senthil:2004aza}
T.~Senthil, \emph{{Deconfined Quantum Critical Points}},
  \href{http://dx.doi.org/10.1126/science.1091806}{\emph{Science} {\bf 303}
  (2004) 1490--1494}.

\bibitem{Senthil:2005jk}
T.~Senthil and M.~P.~A. Fisher, \emph{{Competing orders, non-linear sigma
  models, and topological terms in quantum magnets}},
  \href{http://dx.doi.org/10.1103/PhysRevB.74.064405}{\emph{Phys. Rev.} {\bf
  B74} (2006) 064405}, [\href{http://arxiv.org/abs/cond-mat/0510459}{{\tt
  cond-mat/0510459}}].

\bibitem{Gross:1974jv}
D.~J. Gross and A.~Neveu, \emph{{Dynamical Symmetry Breaking in Asymptotically
  Free Field Theories}},
  \href{http://dx.doi.org/10.1103/PhysRevD.10.3235}{\emph{Phys. Rev.} {\bf D10}
  (1974) 3235}.

\bibitem{Witten:1983ar}
E.~Witten, \emph{{Nonabelian Bosonization in Two-Dimensions}},
  \href{http://dx.doi.org/10.1007/BF01215276}{\emph{Commun. Math. Phys.} {\bf
  92} (1984) 455--472}.

\bibitem{tsvelik_2003}
A.~M. Tsvelik, \emph{Quantum Field Theory in Condensed Matter Physics}.
\newblock Cambridge University Press, 2~ed., 2003,
  \href{http://dx.doi.org/10.1017/CBO9780511615832}{10.1017/CBO9780511615832}.

\bibitem{Affleck:1988zj}
I.~Affleck, \emph{{FIELD THEORY METHODS AND QUANTUM CRITICAL PHENOMENA}},  in
  \emph{{Les Houches Summer School in Theoretical Physics: Fields, Strings,
  Critical Phenomena Les Houches, France, June 28-August 5, 1988}},
  pp.~0563--640, 1988.

\bibitem{Chandrasekharan:1999cm}
S.~Chandrasekharan and U.-J. Wiese, \emph{{Meron cluster solution of a fermion
  sign problem}},
  \href{http://dx.doi.org/10.1103/PhysRevLett.83.3116}{\emph{Phys. Rev. Lett.}
  {\bf 83} (1999) 3116--3119},
  [\href{http://arxiv.org/abs/cond-mat/9902128}{{\tt cond-mat/9902128}}].

\bibitem{Chandrasekharan:2002vk}
S.~Chandrasekharan, J.~Cox, J.~C. Osborn and U.~J. Wiese, \emph{{Meron cluster
  approach to systems of strongly correlated electrons}},
  \href{http://dx.doi.org/10.1016/j.nuclphysb.2003.08.041}{\emph{Nucl. Phys.}
  {\bf B673} (2003) 405--436},
  [\href{http://arxiv.org/abs/cond-mat/0201360}{{\tt cond-mat/0201360}}].

\bibitem{Liu:2019}
H.~Liu, S.~Chandrasekharan and R.~Kaul, \emph{{to be published}}, .

\end{thebibliography}\endgroup

\end{document}